# Vacuum-UV to IR supercontinuum in hydrogen-filled photonic crystal fiber

FEDERICO BELLI[1,*,†], AMIR ABDOLVAND[1,*,†], WONKEUN CHANG[2],
JOHN C. TRAVERS[1] AND PHILIP ST.J. RUSSELL[1]

[1]Max-Planck Institute for the Science of Light, Guenther-Scharowsky Strasse 1, 91058, Erlangen, Germany
[2]Current address: Optical Sciences Group, Research School of Physics and Engineering, The Australian National University, Acton ACT 2601, Australia
† These authors have contributed equally to the work.
*Corresponding authors: federico.belli@mpl.mpg.de; amir.abdolvand@mpl.mpg.de

**ABSTRACT:** Although supercontinuum sources are readily available for the visible and near infrared, and recently also for the mid-IR, many areas of biology, chemistry and physics would benefit greatly from the availability of compact, stable and spectrally bright deep ultraviolet (DUV) and vacuum ultraviolet (VUV) supercontinuum sources. Such sources have however not yet been developed. Here we report the generation of a bright supercontinuum, spanning more than three octaves from 124 nm to beyond 1200 nm, in hydrogen-filled kagomé-style hollow-core photonic crystal fiber (kagomé-PCF). Few-µJ, 30 fs pump pulses at wavelength 805 nm are launched into the fiber, where they undergo self-compression via the Raman-enhanced Kerr effect. Modeling indicates that before reaching a minimum sub-cycle pulse duration of ~1 fs, much less than one period of molecular vibration (8 fs), nonlinear reshaping of the pulse envelope, accentuated by self-steepening and shock formation, creates an ultrashort feature that causes impulsive excitation of long-lived coherent molecular vibrations. These phase-modulate a strong VUV dispersive wave (at 182 nm or 6.8 eV) on the trailing edge of the pulse, further broadening the spectrum into the VUV. The results also show for the first time that kagomé-PCF guides well in the VUV.

## 1. INTRODUCTION

Supercontinuum sources based on solid-core conventional and photonic crystal fiber (PCF) are now a mature technology, providing broadband light from the mid-IR [1] to the near-UV [2,3]. They have not, however, been demonstrated for wavelengths shorter than 280 nm [4], and are found to suffer from limited optical transparency and cumulative optical damage when delivering UV light. As a result no solid-state or fiber-based supercontinuum source exists that can span the VUV (~100-200 nm) and DUV (~200-300 nm) spectral ranges.

Existing sources in the DUV/VUV produce relatively narrow-band light. An example is the use of nonlinear crystals to multiply the frequency of pump laser light [5,6]. Such crystals have, however, limited windows of transparency in the VUV, typically cutting off below ~190 nm (one exception is strontium tetraborate [7]) and in addition they are subject to optical damage. More recently, noble-gas filled kagomé-style hollow-core photonic crystal fiber [8] (kagomé-PCF) has been used to demonstrate widely tunable and efficient generation of ~5 nm wide bands of UV light down to 176 nm [9,10]. Capillaries filled with noble gas are also commonly used for spectral broadening of few-mJ ultrashort pulses [11]. They have been used, for example, to generate light at 270 nm with a bandwidth of 16 nm (corresponding to 8 fs), pumping with a combination of 800 and 400 nm light at pulse energies of ~1 mJ [12].

Supercontinuum generation in bulk gases is widely established [13], and by making use of filamentation has been extended down to the VUV. For instance, extremely flat (but spatially incoherent) supercontinua down to 150 nm were generated via multiple filaments created by focusing TW (250 mJ in 125 fs) laser pulses into noble gases [14]. By using shorter, few cycle pulses, to create a single filament, the required energy was reduced to less than 1 mJ [15], but with a longer UV cut-off wavelength at ~200 nm [16]. The same authors suggested that, if even shorter pulses are used, the supercontinuum may be extended to even shorter wavelengths, but this is yet to be demonstrated.

Here we show how using a hydrogen-filled hollow-core PCF enables generation of a spatially coherent supercontinuum extending from the VUV (124 nm) to the NIR (beyond 1200 nm), using only few µJ, 30 fs pump pulses. This extends, by almost a full octave into the VUV, the previously reported supercontinua generated in free-space, single filaments [16], and exceeds the shortest optical wavelength in a supercontinuum generated to date by any means. It also represents a drastic reduction in the required pulse energy, from the mJ to µJ level, as well as an increase in the required pulse duration, from few cycle pulses to tens of fs. These relaxed pulse source requirements are well within the capabilities of current high power fiber and thin-disk laser technologies [18], opening the possibility for an all-fiber-based VUV-IR supercontinuum source with high average power and MHz repetition rate.

## 2. EXPERIMENTAL DETAILS

### A. Set-up

The experimental set-up (Fig. 1a-b) consists of a 1 kHz amplified Ti:sapphire laser system (Coherent Legend Elite) delivering 30 fs pulses with energies up to a few mJ and a central wavelength of 805 nm. Few µJ pulses were prepared and controlled using an achromatic half-wave plate and a glass wedge at the Brewster angle (not shown in figure). Reflections from a pair of chirped mirrors were used to compensate for the second order dispersion induced by the beam path in air and in the optics. The pulses were launched into a 15 cm long, hydrogen-filled kagomé-PCF and the spectrum emerging from the end-face of the fiber monitored with three different spectrometers. The ends of the kagomé-PCF were mounted in gas-cells, and the whole system filled with hydrogen at uniform pressure. Optical access was provided by 1 mm thick MgF$_2$ windows placed at ~5 cm from the fiber ends and offering a transmission range from the IR down to the VUV with ~81% transmission at 122 nm. Light was coupled into the fundamental mode of the fiber using an achromatic-doublet lens, with launch efficiencies above 80% (defined as the ratio of the output power to the input power just before the fiber input face). The linear coupling efficiency was carefully checked before and after each measurement. Although the fiber end-face could be damaged by accidental misalignment of the launch optics or laser pointing instabilities, no damage was seen within the fiber, which is where the highest intensities occur and the VUV light is generated.

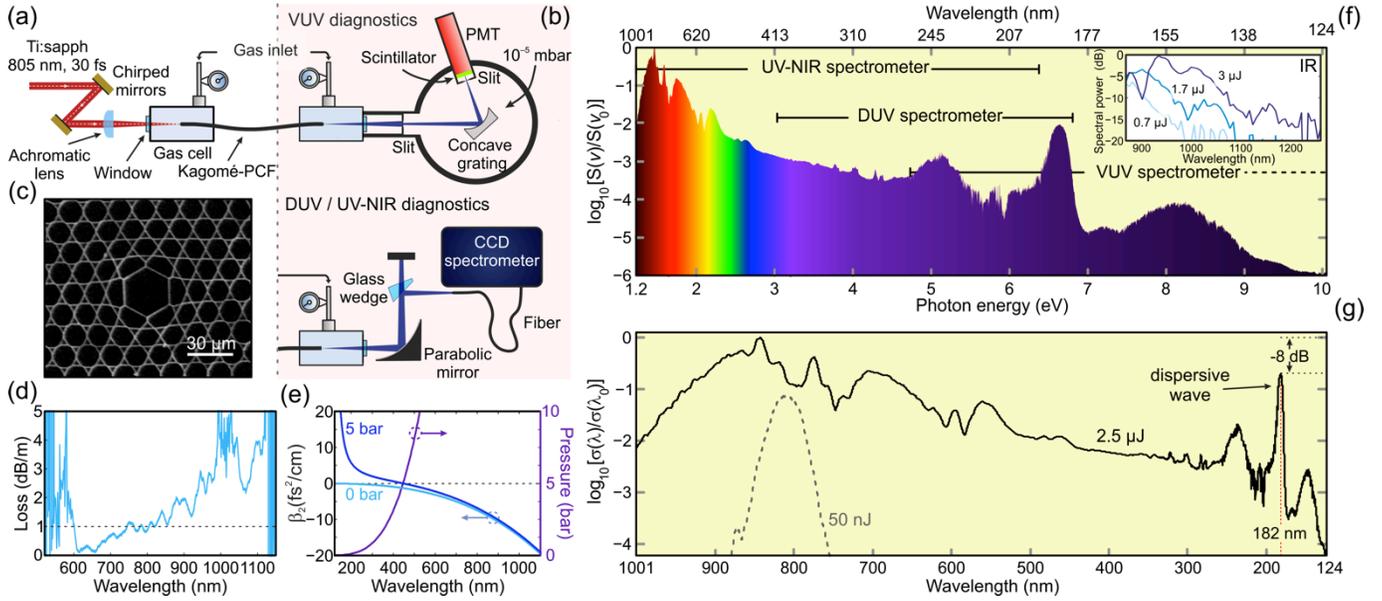

Figure 1: Experimental set-up. (a) A kagomé-PCF is filled with hydrogen using a pair of gas-cells with MgF$_2$ windows. Ultra-short few μJ pulses are launched into the fiber using an achromatic lens. Two chirped mirrors compensate for pulse lengthening in air. (b) Diagnostics. Upper: the VUV spectrum is measured with an evacuated scanning monochromator equipped with a scintillator and a photomultiplier tube (PMT). Lower: The UV-NIR and DUV spectra are measured by directing the beam to either a UV-NIR or a DUV CCD-based spectrometer using a UV-enhanced optical fiber and a parabolic mirror. Before entering the spectrometer fiber, the signal is attenuated by multiple reflections from several glass wedges. (c) Scanning electron micrograph of the cross-section of the kagomé-PCF. (d) Experimentally measured loss curve for the kagomé-PCF used in the experiments. (e) Left-hand axis: Calculated group velocity dispersion $\beta_2$ plotted against wavelength (lower axis) for an evacuated fiber (light blue) and a fiber filled with hydrogen at 5 bar pressure (dark blue). Right-hand axis: hydrogen pressure needed to produce a given zero-dispersion wavelength (lower axis). (f) Experiment: Supercontinuum spectrum generated by pulses of duration 30 fs, centre-wavelength 805 nm and energy 2.5 μJ after propagation along 15 cm of kagomé-PCF filled with hydrogen at 5 bar. The spectra were obtained using three different spectrometers as indicated. The frequency-scaled spectral energy density $S(\nu)$ (obtained from the measured wavelength-scaled spectral energy density $\sigma(\lambda)$ through $S(\nu) = \sigma(\lambda) \lambda^2/c$) is normalized to its value at the peak of the spectrum (1.45 eV). The strong peak at ~6.8 eV (182 nm) is caused by dispersive wave generation [17]. Inset: Supercontinuum spectra in the IR for increasing launched pulse energy (marked on plot), measured using an uncalibrated fiber-based spectrometer. (g) Spectral energy density in (f) recalibrated in terms of wavelength $\sigma(\lambda)$ (solid black line). Note a drop of only 8 dB for the dispersive wave peak at 182 nm compared to the pump at 805 nm. The dashed grey line shows the initial spectrum of the pump as measured at 50 nJ and before experiencing extreme spectral braodening.

### B. Choice of nonlinear medium

Among the different molecular gases, hydrogen offers several advantages: a broad transmission range down to the extreme UV (~ 70 nm), a long dephasing time $T_2$ of its molecular oscillations (1 ns and 3 ns, respectively for rotational and vibrational modes at 1 bar), a large high-pressure Raman gain coefficient (e.g. 1.55 cm/GW for vibrational mode [19]), a relatively high ionization potential (15.43 eV) and the highest rotational and vibrational frequency shifts of all molecules, with values of $\Omega_{rot}/2\pi = 17.6$ THz (pure rotational transition, S(1)) and $\Omega_{vib}/2\pi = 124.56$ THz (pure vibrational transition, Q(1)) for ortho-hydrogen. As will be shown, the combination of these factors is essential in achieving the observed super-broadening in H$_2$.

### C. Choice of waveguide

Two unique properties of the kagomé-PCF are vital to the success of the experiment: (i) high ultrabroad transmission (even, as we show here, in the VUV) for core diameters below 40 μm (ii) pressure-tunable modal properties, providing anomalous dispersion at the pump wavelength over a considerable range of gas pressures.

It has been experimentally and numerically verified [17,20] that the wavelength dependence of the modal refractive index in a gas-filled kagomé-PCF in the visible and UV regions can be approximated to high accuracy by a simple hollow capillary, taking the form [21]:

$$n_{nm}(\lambda) = [n_{gas}^2(\lambda) - (u_{nm}\lambda/(2\pi a))^2]^{1/2} \qquad (1)$$

where $n_{gas}$ is the refractive index of the gas, obtained from a Sellmeier expansion [22], $a$ is the core radius, $u_{nm}$ is $n$th zero of the $(m-1)$th order Bessel function of the first kind, associated with the HE$_{nm}$ hybrid mode of the hollow waveguide.

The kagomé-PCF used in the experiments (Fig. 1c) was fabricated using the stack-and-draw technique. It had a cladding period of 13.4 μm, a core diameter of 25 μm and a core-wall thickness of ~150 nm, estimated from the scanning electron micrograph in Fig. 1c. The loss spectrum (Fig. 1d, measured by the cut-back technique) in the NIR and VIS ranges was ~1 dB.m$^{-1}$, reaching a minimum value of 0.1 dB.m$^{-1}$ at 658 nm. Plots of the calculated modal dispersion (using Eq. 1) for the evacuated fiber, and one filled with 5 bar of hydrogen, are shown in Fig. 1e. The dispersion at the pump wavelength is anomalous for all the pressures used; this is of vital importance in the experiments. Although a simple capillary of the same bore diameter offers the same dispersion, its transmission losses would be prohibitively high at ~600 dB.m$^{-1}$ at 800 nm.

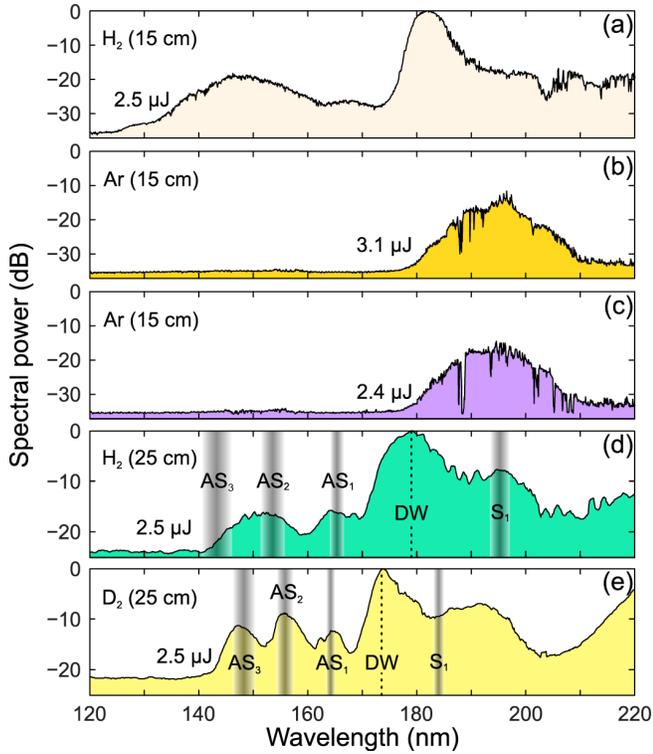

Figure 2: Experimental VUV spectra recorded when a 15 cm long kagomé-PCF was filled with (a) hydrogen at 5 bar and pumped with with 2.5 µJ pulses and (b, c) argon at 4 bar and pumped with pulses of energies 2.4 µJ and 3.1 µJ. (d&e) The experimental VUV spectra recorded for a 25 cm long kagomé-PCF filled with (d) hydrogen and (e) deuterium at 5 bar and pumped with 2.5 µJ pulses. The dashed lines show the position of the dispersive wave (DW, black) in each case. The shaded bars, corresponding to manifolds of rotational-vibrational transitions, mark the expected positions of the first Stokes ($S_1$) and higher-order anti-Stokes lines ($AS_n$, n=1,2,3, blue) for a pump at the dispersive wave position.

### D. Data collection and analysis

The light emerging from the fiber end-face was monitored either using an evacuated VUV spectrometer (McPherson) directly connected to the output gas-cell, or by attenuating the signal by reflection at one or more glass wedges and then coupling it into different fiber-based CCD spectrometers using an off-axis parabolic mirror (Fig. 1b). The VUV spectrometer was equipped with a scintillator and photomultiplier tube (PMT) and was initially purged and then evacuated to a pressure of 10$^{-5}$ mbar so as to prevent absorption by air. The spectral response of the VUV spectrometer was carefully calibrated using the known grating diffraction efficiency, the known quantum efficiency of the scintillator detector and by estimating the divergence of the light emitted from the fiber end. The spectra obtained from the three different spectrometers were combined by rescaling their overall spectral intensities so as to match in the overlap regions.

## 3. EXPERIMENTAL RESULTS

### A. Hydrogen

Figures 1f and 1g show respectively in the frequency and wavelength domains, the spectrum generated in a 15 cm length of fiber filled with hydrogen at 5 bar and pumped by pulses of 2.5 µJ. At this pressure the zero dispersion point is at 446 nm (Fig. 1e). The supercontinuum spans from 124 to beyond 1200 nm – measurements with a mid-IR spectrometer (with uncalibrated absolute response) showed a strong signal from 1000 to 1200 nm (inset of Fig. 1f). To check that the supercontinuum was not the result of averaging over many shots, a single-shot measurement was made in the IR spectral region. This entailed an integration time of 1 ms (the laser repetition rate was 1 kHz), which limited the dynamic range of the spectrometer to 20 dB at best. The spectrum is also impaired by a band of very high fiber loss beyond 1.15 µm (see Fig. 1d). This is the first report to our knowledge of a supercontinuum that extends into the deep and vacuum UV.

No degradation in the coupling efficiency, the transmitted spectral shape or the output power was observed even after weeks of lengthy experiments using the same piece of fiber. This robustness to damage at wavelengths well below 200 nm in the VUV, can be ascribed to the negligible overlap between the guided core mode and the silica core-wall and cladding structure [23]. This has the effect of drastically reducing glass solarisation by DUV-VUV light (the electronic band-edge of silica is at ~160 nm), and even if the glass is damaged, the effect on the guided light is marginal because the light-glass overlap is so small. This absence of optical damage at such short wavelengths makes the system a remarkable advance over sources based on solid-state materials.

### B. Comparison between hydrogen and argon

To clarify the role of the Raman contribution in extending the supercontinuum spectrum into the VUV, we directly compared spectral broadening in H$_2$ and Ar. Argon was selected as it has very similar properties to H$_2$ in terms of dispersion, ionization potential and Kerr nonlinearity, however, it is monoatomic and therefore possesses no Raman nonlinearity, making it a perfect candidate for isolating the role of the Raman contribution in the spectral broadening.

Three factors are of importance to the nonlinear dynamics: the dispersion landscape, the nonlinearity experienced by the pulse, and the ionization potential, which is very similar for Ar (15.76 eV) and hydrogen (15.43 eV). Following Ref. [8,17], for fixed input pulse duration and wavelength, the Kerr-based nonlinear dynamics can be parameterised by the zero dispersion wavelength (ZDW), and the soliton order, defined as $N = \sqrt{\gamma P_0 \tau_p^2 / |\beta_2|}$ where $\gamma$ is the nonlinear coefficient (directly proportional to $\chi^{(3)}$), $P_0$ is the peak power, $\tau_p$ is the pulse duration and $\beta_2$ the group velocity dispersion [17]. It turns out that in the case of Ar and H$_2$ these two parameters can be matched simply by adjusting the gas-filling pressure and input pulse energy. The ZDW is at 446 nm for 5 bar of H$_2$ and 4 bar of Ar. At these pressures the soliton order is ~5 for input energies of 2.5 µJ (hydrogen) and 2 µJ (argon).

Given that Ar has slightly higher values of $\chi^{(3)}$ and ionization potential, and that the pressure is lower, one might expect that (in the absence of contributions from the Raman effect) the supercontinuum generated in Ar should extend further into the VUV [24].

The experimental results tell a very different story, however. Fig. 2a-c compares the observed short-wavelength spectra generated in hydrogen (Fig. 2a) with that generated in argon (Fig 2b-c). Note that the pulse energy in the Ar case was scanned up to 3.1 µJ so as to be sure to reach the same effective nonlinearity as in the H$_2$ experiment, where the pulse energy was 2.5 µJ. It is clear from the measurements that, when the dispersion landscape and effective nonlinearity are matched, the H$_2$-supercontinuum extends much further into the VUV. The strongest VUV peaks, which correspond to dispersive wave (DW) emission into the fundamental HE$_{11}$ mode [9,10], are centered at 182 nm for H$_2$ (black) and 195 nm for Ar. These spectral positions approximately follow a simple phase-matching condition [17,25].

### C. Comparison between hydrogen and deuterium

To further clarify the mechanism by which the Raman contribution drives the spectral extension into the VUV, we directly compared spectral broadening in H$_2$ and D$_2$ using a 25 cm long kagomé-PCF (Figs. 2d-e). Note that the longer fiber length (c.f. Fig. 2a) introduces additional transmission loss, particularly at the shortest wavelengths. Since the dispersive waves and the third and fifth harmonics have smaller group velocities than the rest of the spectrum, they lag behind the compressed pulse and so are strongly phase-modulated by the oscillating refractive index of the vibrational coherence wave [26] – see Sec. 4. This effect would be enhanced over longer propagation lengths provided phase-matching is maintained. The shaded regions in Fig. 2d and 2e mark the expected positions of the first Stokes (S$_1$) and higher-order anti-Stokes lines AS$_n$ (relative to the DW frequency), corresponding to a manifold of rotational-vibrational transitions of the form $|v,J\rangle \to |v',J'\rangle$ with $|\Delta v| = |v - v'| = 1$ and $|\Delta J| = |J - J'| = 0, 2$, where $v$ and $J$ are the vibrational and rotational quantum numbers. Elevated values of the spectral power are seen in these regions for both H$_2$ (Fig. 2d) and D$_2$ (Fig. 2e). Note that the fine details of these transitions on both Stokes and anti-Stokes sides could not be resolved experimentally due to the broad spectrum of the driving DW.

## 4. DISCUSSION

Most gas-based studies of supercontinuum generation to date have focused on atomic gases, the spectral broadening being mainly attributed to self-phase modulation (SPM) and self-steepening [13,27]. In this case the intensity dependent change of the refractive index, caused by the (non-resonant) instantaneous Kerr nonlinearity, results in SPM and spectral broadening of the pulse. The rather unconventional choice of hydrogen – a Raman active rather than a noble gas – leads to richer but more complex nonlinear dynamics. This is mostly related to the highly non-instantaneous response of molecular gases to an external excitation that has temporal features that are faster than the period of molecular oscillation [28]. Under these circumstances the Raman transitions are driven by frequencies higher than their natural frequencies, which has the additional consequence that, unlike the Kerr effect, the Raman-induced refractive index change can be both positive and negative, causing a strong modulation of the instantaneous frequency and temporal profile of the pulse.

### A. Raman response in different regimes

The Raman response depends on the pump pulse duration relative to the period of molecular oscillation, $T_m = 2\pi/\Omega$ and the dephasing time of molecular oscillations, $T_2$. Before considering propagation dynamics, we briefly summarize the characteristics of the different Raman regimes.

Pumping a hydrogen-filled kagomé-PCF with long, narrow-band, high-energy laser pulses results in the generation of a cascade of discrete vibrational and rotational Stokes and anti-Stokes Raman side-bands [29]. If two pump lasers are used, detuned so that their frequency difference is close to a Raman transition, i.e. $\omega_P - \omega_S \simeq \Omega$ (where subscripts P and S refer to pump and Stokes), then the molecular coherence can be driven [28–31]. The ensemble of driven molecules then acts back on the light, resulting in the generation of a discrete 4 comb spaced by the Raman frequency shift $\Omega$. In this case, the two driving pulse durations, $\tau_P$ and $\tau_S$, must be such that $\tau_P \sim \tau_S \gg T_m$ (usually nanosecond pulses are used in the experiment) so that in the frequency domain one observes the appearance of spectrally well-separated discrete Raman side-bands [30]. While this approach has been used to generate frequency combs spanning from the IR to VUV [32,33], it is not suitable for producing a broadband supercontinuum.

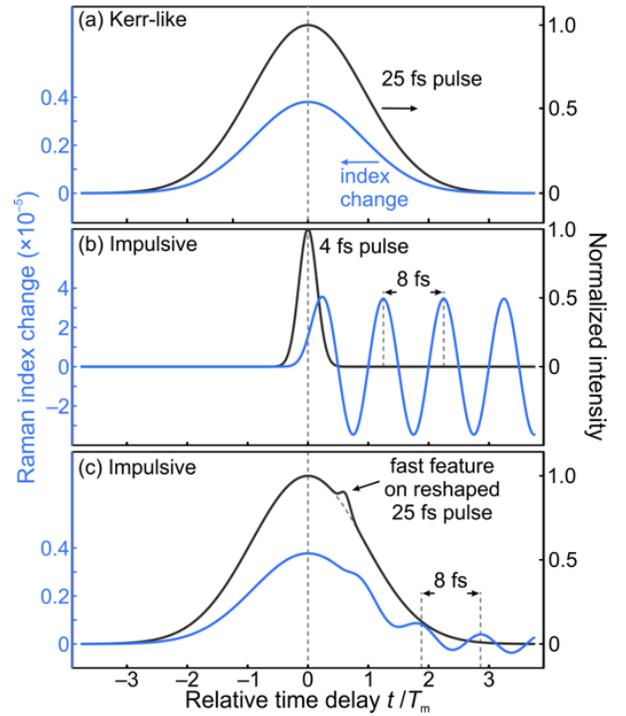

Figure 3: Illustrating the different regimes of Raman scattering. The pulse intensity profiles (right-hand axis) and Raman index modulation (left-hand axis) are plotted against time delay in units of the Raman oscillation period $T_m = 8$ fs. (a) Instantaneous Kerr-like response when the pulse duration (25 fs) is much longer than the Raman oscillation period. (b) Impulsive Raman scattering when a very short (4 fs) pulse impinges on the gas, exciting a strong Raman-related index oscillation at 125 THz. The positive index slope under the pulse red-shifts its frequency. (c) Raman oscillations created when a long (25 fs) pulse is reshaped by self-steepening, resulting in a very fast feature (indicated by the arrow) that is able to impulsively drive the Raman oscillations.

In Fig. 3 we illustrate the different regimes of Raman responses associated with different pump pulse durations for the case of vibrational Raman excitation of H$_2$ ($T_m = 8$ fs). In the so-called transient regime $T_m \ll \tau_P < T_2$, for a pulse of duration 25 fs (Fig. 3a), the induced refractive index change (blue) follows the intensity profile of the long – compared to $T_m$ – exciting pulse (black). The molecule undergoes many cycles of oscillation over the duration of the pulse. This regime is similar to the case of a Kerr material, where the electronic response is faster than the exciting pulse.

The situation changes when a single, close to transform-limited, pump pulse, with a bandwidth greater than the Raman frequency shift of the medium, impulsively excites coherent molecular oscillations (Fig. 3b). In the temporal domain this means that the pulse duration is shorter than both the Raman oscillation period and the dephasing time, i.e., $\tau_P < T_m \ll T_2$. This regime is known to induce continuous spectral broadening of an ultrashort pulse, accompanied by a redshift in its central wavelength that is caused by the Raman contribution to the refractive index (Fig. 3b), which increases with time within the pulse [26,34]. This regime is reminiscent of the soliton self-frequency shift well-known in solid-core fiber, but with substantial differences. For example, the Raman oscillations in hydrogen have a dephasing time that, depending on the pressure, ranges from several ns to a few hundreds of fs [35]. This is orders of magnitude longer than in silica, where the dephasing time is only a few fs.

In Fig. 3c we depict the case of a pulse that has been reshaped such that it has a temporal feature short enough to impulsively drive Raman oscillations, while the overall pulse duration is longer. This has sometimes been referred to as the "displacive" regime [36,37]. Its main difference from (and advantage over) the impulsive regime (Fig. 3b) is that the overall pulse duration is still long enough for its trailing edge to feel the effects of the strong self-induced refractive index oscillation and be shifted to both higher and lower frequencies. This regime thus plays a key role in the VUV spectral broadening reported here.

We note here that the timescales in the experiment (<30 fs) are much shorter than the timescale for full revival of the rotational wavepackets, which is ~250 fs for $H_2$ and ~500 fs for $D_2$ (given by $(2cB)^{-1}$ where $B$ in m$^{-1}$ is the rotational constant and $c$ is the speed of light [38]). As a result orientational effects do not play a significant role. Induced birefringence in the gas and the revival of rotational wave-packets after the passage of the pulse should however be easily detectable in a pump-probe scheme.

### B. Numerical modeling

We modeled the propagation of an ultrashort pulse in hydrogen-filled kagomé-PCF by numerically solving Maxwell's equations. The Raman polarization was included via a set of Maxwell-Bloch equations for the off-diagonal (related to coherence) and diagonal (related to population) elements of a 2×2 density matrix. As the pulse durations are extremely short, numerical models based on the slowly varying envelope approximation (SVEA) are not reliable. Therefore, in order to account for the evolution of both the full electric field and the associated nonlinear material response we used a multi-mode formulation [39], including both spatial and temporal effects, of a unidirectional field equation [40] that is closely related to the well-known unidirectional pulse propagation equation (UPPE) [41].

The scalar nonlinear polarization consists of three terms: $P_{NL}(r,z,t) = P_K + P_R + P_e$ where $P_K = \varepsilon_0 \chi^{(3)} E^3(r,z,t)$ is the instantaneous part related to the Kerr effect ($\chi^{(3)}$ for hydrogen is $2.206 \times 10^{-26}$ m$^2$.V$^{-2}$ under standard conditions [42]), where $E(r,z,t)$ is the radially symmetric real-valued electric field, $P_R$ is related to the Raman effect, and $P_e$ is induced by photoionization of the gas and subsequent evolution of free electrons in the electric field. These last two terms are discussed in the following subsections.

#### 1. Raman polarization

In order to account for the effect of the different molecular degrees of freedom we assume that the rotational and vibrational Raman modes are decoupled. This leads to two independent contributions to the Raman polarization $P_R = P_R^{(rot)} + P_R^{(vib)}$ where

$$P_R^{(k)} = N_g \text{Tr}\left[\hat{\rho}^{(k)} \hat{\alpha}^{(k)}\right]$$
$$= N_g \left[\alpha_{11}^{(k)} + \left(\alpha_{22}^{(k)} - \alpha_{11}^{(k)}\right)\rho_{22}^{(k)} + 2\alpha_{12}^{(k)} \text{Re}\left(\rho_{12}^{(k)}\right)\right]E, \quad \textbf{(2)}$$

$\hat{\rho}^{(k)}$ and $\hat{\alpha}^{(k)}$ are the 2×2 density and polarizability tensors, with matrix elements $\alpha_{ij}$ and $\rho_{ij}$, associated with the rotational (k = rot) S(1) Raman branch of ortho-hydrogen, and the vibrational (k = vib) Q(1) branch. $N_g(p, T)$ is the molecular number density of the gas at pressure $p$ and temperature $T$.

Under the assumption of a single-photon resonance far-detuned from the laser frequency [30,43,44], the induced nonlinear Raman polarization and the population inversion can be described by the following pair of equations, without using the SVEA:

$$\left(\partial_t + 1/T_2^{(k)} - i\Omega_k\right)\rho_{12}^{(k)}$$
$$= \frac{i}{2\hbar}\left[\left(\alpha_{11}^{(k)} - \alpha_{22}^{(k)}\right)\rho_{12}^{(k)} + \alpha_{12}^{(k)} w^{(k)}\right]E^2 \quad \textbf{(3)}$$

$$\partial_t w^{(k)} + \frac{w^{(k)} + 1}{T_1^{(k)}} = -\frac{2\alpha_{12}^{(k)}}{\hbar} \text{Im}\left\{\rho_{12}^{(k)}\right\} E^2 \quad \textbf{(4)}$$

where $w^{(k)} = \rho_{22}^{(k)} - \rho_{11}^{(k)}$ and $\rho_{11}^{(k)} + \rho_{22}^{(k)} = 1$. In these equations $T_1^{(k)}$ is the lifetime of the excited rotational and vibrational levels (taken to be ~20 ns in both cases), and $T_2^{(rot)} = \frac{318.3}{6.15/\eta + 114\eta}$ ns and $T_2^{(vib)} = \frac{318.3}{309/\eta + 52\eta}$ ns are the coherence lifetimes and $\eta = N_g / N_{g0}$, where $N_{g0}$ is the gas number density in standard conditions [35]. The values of the polarizability tensor elements used in the simulations are $(\alpha_{11}, \alpha_{22}, \alpha_{12}, \alpha_{21}) = (8.9, 9.4, 0.85, 0.85) \times 10^{-41}$ F m$^2$ for rotation and $(\alpha_{11}, \alpha_{22}, \alpha_{12}, \alpha_{21}) = (8.9, 9.73, 1.55, 1.55) \times 10^{-41}$ F m$^2$ for vibration. In calculating the polarizability tensor elements we have carefully taken into account the average over different orientations of the molecular axis with respect to the direction of the electric field. Note that these values are smaller than those reported in [44,45], but have been carefully checked against *ab initio* calculations and experimental data.

#### 2. Photoionization and plasma evolution

Following Ref. [39–41] we model the nonlinear polarization arising from photoionization and the evolution of free electrons using the photoionization rate model in [46]. The theory of photoionization in molecular gases is at present an active research area [47], and still not fully resolved. When the experimental ionization potential of 15.43 eV is used, the tunnelling-based Ammosov-Delone-Krainov model [48] (ADK) overestimates the ionization rate, compared to full *ab initio* calculations [49]. Since we are not strictly operating in the tunnel-ionization regime, in all the simulations we used the Perelomov-Popov-Terent'ev (PPT) model [50], which includes multi-photon ionization and also overestimates the ionization rate. Even so, numerical simulations show that plasma formation does not strongly contribute to the propagation dynamics for the parameter regimes explored here.

### C. Propagation dynamics

Figure 4 shows the modeled spectral (Fig. 4a) and temporal (Fig. 4d-4g) evolution of a 2.5 µJ, 30 fs pulse launched into the kagomé-PCF filled with hydrogen to a pressure of 5 bar. These parameters match the values in the experiment. A detailed comparison between theory and experiment at 1.7 µJ is shown in Fig. 4b along with the experimentally and numerically observed spectral evolution as a function of the launched pump energy in Fig. 4c.

The numerical simulations show that the pulse duration at the point of maximum compression is 1.3 fs (Fig. 4f-4g). This dramatic soliton-effect compression is the result of the broad and low anomalous dispersion offered by the fiber around the pump wavelength, even when filled with normally-dispersive gas at relatively high pressure.

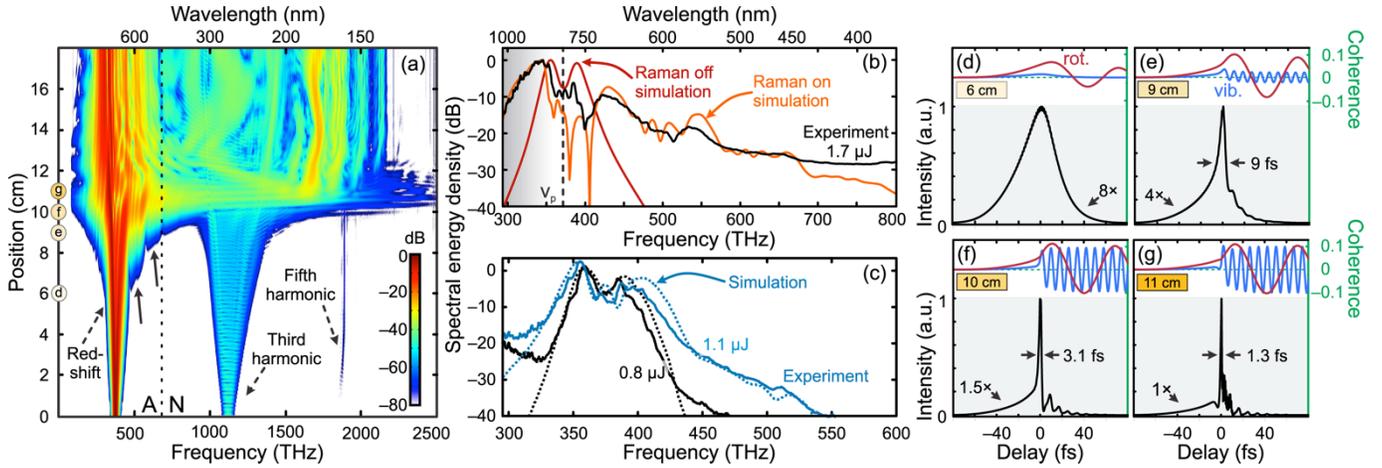

Figure 4: Numerical simulations and experiment. (a) Calculated spectral evolution along the fiber at 5 bar hydrogen for input pulse energy 2.5 µJ and duration 30 fs. At ~10 cm the pulse has compressed to a duration of ~3 fs. Third and fifth harmonics are also visible. The dotted vertical line marks the position of the zero dispersion wavelength (446 nm). The dashed arrow on the left indicates the initial red-shift due to impulsive rotational Raman modulation, while the solid arrows indicate the expected positions of first and second vibrational anti-Stokes lines of a narrow-band 805 nm pump. (b) Comparison between experiment and simulation at 1.7 µJ. When the Raman contribution is turned off the simulations fail to predict the observations. (c) Experimentally observed spectral broadening of the pump pulse with increasing pump pulse energy (solid lines) and the corresponding simulated results (dotted lines) show very good agreement. (d-g) Numerical simulations of the envelope of the optical intensity (black line) together with the rotational (red) and vibrational (blue) coherence waves at four different positions along the propagation. The corresponding positions are marked on the vertical axis of the propagation plot in (a).

Soliton-effect compression is well known to occur in kagomé-PCF filled with a Raman-free noble-gas such as argon [9,10,17] and is usually characterized by a smooth tail toward higher frequencies [51]. In hydrogen-filled fiber, however, Raman-related nonlinearities play an equally important role. Even before self-compression sets in, rotational oscillations (period 56 fs) are impulsively driven by the $\tau_p = 30$ fs pump pulse (see Fig. 4d). This results in a Raman-induced rotational index change that increases with time under the pulse, causing it to red-shift in frequency ($\Delta\omega(t) \propto -\partial n/\partial t$), which explains the spectral enhancement at lower frequencies observed in the experimental results (grey-shaded area in Fig. 4b). The total Raman-related index change, including both rotational and vibrational responses, strengthens SPM beyond that provided by the electronic Kerr effect alone [52], enhancing self-compression [34] and self-steepening. Note that, since $\tau_p \gg 8$ fs, the vibrational Raman response "instantaneously" follows the intensity profile, i.e., it is Kerr-like.

In the experiment, a strong UV spectral extension develops at higher input energies (Fig. 4b-4c), which disappears when the Raman effect is switched off in the simulations (Fig. 4b). In broad outline, this dramatic broadening into the VUV arises as follows. Raman-enhanced self-steepening causes a very sharp temporal feature to develop on the trailing edge of the pulse. This generates a broad spectrum (see below), which enhances the transfer of energy to a dispersive wave in the VUV (6.8 eV, 182 nm), beyond what is seen when noble gases are used [9,10]. It is also short enough to cause strong impulsive excitation of vibrational coherence, which then phase-modulates the residual trailing edge of the pulse, including the dispersive wave, creating new bands of frequency in the DUV and VUV. This neatly explains the results in Figs. 2a-c, where the Raman effect was switched off in the experiments by replacing hydrogen with argon, resulting in the disappearance of the VUV extension, and also the results in Figs. 2d-e, where replacing hydrogen with deuterium results in the appearance of DW sidebands at frequencies corresponding to the smaller Raman shift of $D_2$.

Taking a closer look at the simulated spectral (Fig. 4a) and temporal (Fig. 4d-4g) behaviour, we see that initially (for distances $z < 7$ cm) the pulse is much longer than 8 fs, and the vibrationally-induced refractive index change follows the intensity envelope (blue solid line in Fig. 4d). Upon further propagation the trailing edge of the pulse gradually self-steepens, until between 7 and 9 cm it has developed a feature short enough to impulsively excite vibrational coherence (Fig. 4e). The spectral features at ~498 THz and ~623 THz (marked by solid arrows in Fig. 4a) lie on the first two vibrational anti-Stokes lines of the 373 THz pump, further confirming the presence of vibrational coherence.

As the pulse propagates further, a succession of similar but gradually weaker events takes place, spaced by the Raman period, further amplifying the vibrational coherence (Fig. 4e-4g) [53]. At z ~ 10 cm (Fig. 4f) even the most intense part of the pulse is short enough to impulsively excite vibrational coherence. Indeed at the maximum compression point (z ~ 11 cm) the spectrum extends down to 2500 THz (~120 nm). This is however a transient feature (a shock) that is rapidly followed by nonlinear spectral narrowing upon further propagation, reducing the high frequency edge of the spectrum to below ~2200 THz.

Third harmonic generation at 268.3 nm, shown by a dashed arrow in Fig. 4a, also plays a role in the evolution of the spectrum, because it overlaps spectrally with the 6th vibrational anti-Stokes band at 267.7 nm. Seeding a Raman process with harmonics of the pump that match an anti-Stokes line is known to push the spectrum towards higher frequencies [32]. This effect is further enhanced at 5 bar pressure, when the pump and third harmonic share similar group velocities.

Note that owing to computational restrictions the simulations do not include resonantly enhanced Raman polarizabilities, which can increase by almost an order of magnitude at VUV wavelengths. Moreover, as the VUV extension of the supercontinuum approaches electronic resonances, the validity of the molecular polarizability model, based on the assumption of off-resonance Raman excitation, breaks down. We believe these factors explain why the simulations presented in Fig. 4b underestimate the spectral energy density for the high-frequency edge of the supercontinuum (above 700 THz).

Small fluctuations in pulse duration, chirp or energy will alter the onset of the broadening. When averaged over several laser shots in the experiment (with the shortest integration time, the UV-NIR spectrometer captured a sequence of 13 pulses), the depths of the spectral dips near the pump frequency are reduced compared to single-shot simulations (Fig. 4b).

## 5. SUMMARY

The fiber-based supercontinuum source reported here represents a breakthrough in the generation of broad-band DUV-VUV light, filling an availability gap in a spectral region very important in spectroscopy [54–56]. By integrating the spectrum, calibrated using best estimates of the instrument response, we estimate a conversion efficiency of 5% into the VUV. This is higher than most other schemes [7], although it will require rigorous verification using calibrated detectors. Due to its broad bandwidth, the spectral brightness of this very compact table-top source is relatively low compared to expensive large-scale synchrotrons. This can be readily improved, as the pump pulse requirements can be met using high repetition rate (MHz) laser sources (e.g., fiber lasers), providing a straightforward route to much higher spectral energy densities. However, even at 1 kHz the brightness is already more than sufficient for single-photon techniques such as photoemission spectroscopy.

Since the light is produced in a gas-filled kagomé-PCF, it may also be flexibly delivered by the same fiber, neatly avoiding the problem of absorption in air. Additionally, it may also be possible to compress the supercontinuum spectrum to produce ultrashort pulses – the bandwidth required for a 1 fs transform-limited Gaussian pulse at 180 nm is 440 THz, equivalent to a bandwidth of 48 nm. Such short VUV pulses would enhance the temporal resolution available in fields such as two-dimensional ultrafast spectroscopy and analytical chemistry [57,58]. The results add further support to Michael Downer's 2002 prediction that gas-filled hollow core PCF would launch "a new era in the nonlinear optics of gases, and maybe even plasmas" [59].